# Advances in Reversible Covalent Kinase Inhibitors

Zheng Zhao[1,2]* and Philip E. Bourne[1,2]*

[1.]School of Data Science and [2.]Department of Biomedical Engineering, University of Virginia, Charlottesville, Virginia 22904, United States of America

*Corresponding author

Email: zz7r@virginia.edu (Z.Z.) and peb6a@virginia.edu (P.E.B.)

## Table of contents graphic

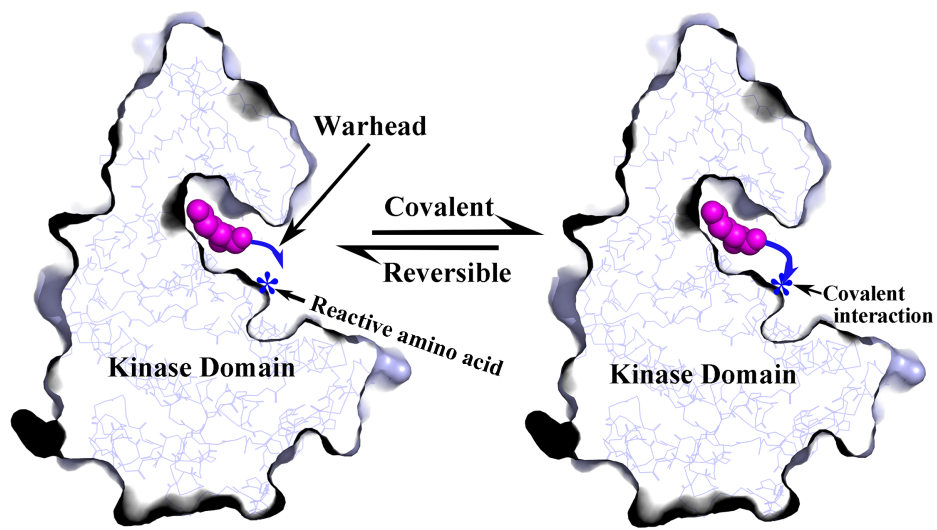

# Abstract

Reversible covalent kinase inhibitors (RCKIs) are a class of novel kinase inhibitors attracting increasing attention because they simultaneously show the selectivity of covalent kinase inhibitors, yet avoid permanent protein-modification-induced adverse effects. Over the last decade, RCKIs have been reported to target different kinases, including the Atypical group of kinases. Currently, three RCKIs are undergoing clinical trials. Here, advances in RCKIs are reviewed to systematically summarize characteristics of electrophilic groups, chemical scaffolds, nucleophilic residues, and binding modes. In so doing, we integrate key insights into privileged electrophiles, the distribution of nucleophiles, and hence effective design strategies for the development of RCKIs. Finally, we provide a further perspective on future design strategies for RCKIs, including those that target proteins other than kinases.

# 1. Introduction

Kinases are one of the largest protein families in the human genome, comprising 538 kinase-encoding genes[1] which catalyze the transfer of the γ-phosphate of ATP to phosphorylate specific substrates[2]. In so doing, kinases mediate cell signal transduction which controls a variety of biological functions such as cell proliferation and apoptosis[3]. Not surprisingly, mutations and dysfunction of kinases are associated with a variety of disease conditions such as cancer, inflammatory disease, cardiovascular disease, neurodegenerative disease, and metabolic disease[4]. As such, kinases have become important therapeutic targets[5-8], however, due to the high conservation of the ATP binding site, it is a challenge to design drugs that target kinases with the desired selectivity[9-10].

As structural and functional knowledge of kinases has increased, substantial variations – such as allosteric sites – have been found among different kinases[11], allowing for greater specificity. Since the first kinase drug, Imatinib, was approved by the U.S. Food and Drug Administration (FDA) in 2001, kinase-targeted drug discovery has been one of the fastest-growing areas of drug development[10, 12]. As of May 29, 2021, 71 small-molecule kinase drugs have been approved by the FDA[13] and over 175 kinase inhibitors are in clinical trials[14]. Correspondingly, there are over 300 kinase targets with released crystal structures providing a substantial structural basis for kinase drug discovery[15-16]. Undoubtedly, kinase-targeted drug discovery has significantly contributed to clinical targeted therapy, for example, against multiple types of cancer, such as non-small cell lung cancer (NSCLC), melanoma, and leukemia[17]. However, in practice, toxicity and adverse events such as congestive heart failure and cardiogenic shock in some chronic myelogenous leukemia patients[18], require the further development of novel, effective and safe inhibitors[9, 19].

Of the 71 currently marketed drugs, 63 are non-covalent inhibitors characterized as Type-I, II, or III according to their binding modes[7, 20]. The remaining 8 are irreversible-covalent kinase inhibitors that utilize a non-catalytic cysteine within or near the ATP binding site forming the covalent interactions[21]. Although irreversible covalent kinase inhibitors can have improved selectivity, concerns exist regarding the potential toxicities of such irreversible complexes[22-24]. Recently, different schemes to reduce the potential toxicities have been provided, notable here are limited reactivity irreversible covalent inhibitors[25-26]. This type of inhibitor is generally considered to be a substrate. The substrate is first catalyzed by an enzyme, leading to the formation of a long-lasting covalent intermediate able to achieve the desired inhibitory effect. Then with the catalytic action of the enzyme, the covalent bond is broken, the active enzyme is released again and the inhibitor (substrate) is catalyzed into other products. Currently, the application of this strategy to design low-reactive irreversible covalent inhibitors is just used with enzymes having covalent catalytic mechanisms[26]. However, because kinase inhibitory mechanisms don't involve covalent interactions with the substrate, the "limited-reactivity" strategy is not suited to developing covalent kinase inhibitors. Another reversible-covalent approach involves an inhibitor with one electrophile first binding the corresponding binding site, then the electrophile reacts reversibly with one nucleophile from an amino acid side chain[27-30]. The reversible covalent binding mode not only ensures a high potency as in covalent interactions, but also allows the adjustment of the residence time by tailoring the electrophilic group. Therefore, reversible-covalent inhibitory mechanism has been applied in developing effective kinase inhibitors[27-31], aptly named reversible-covalent kinase inhibitors (RCKIs) [27, 29]. To date a variety of RCKIs have been reported; indeed, three have been tested in clinical trials[32-34]. Here we provide a systematic review of current progress with RCKIs and provide a look ahead. We first focus on all the reported RCKIs and then describe the intrinsic properties of their reversible-covalent reactive warheads, their binding modes, and binding site contextual information. From there we speculate on privileged warheads, nucleophilic groups, and design strategies for future RCKIs.

## 2. Overview of RCKIs

### 2.1 Kinase inhibitors

Various kinase inhibitors, such as Type-I, Type-II, Type-III, and Type-IV, have been developed to achieve the desired selectivity by making full use of the different features of the ATP binding site and beyond[9, 12]. Type-I inhibitors, such as Crizotinib, (Figure 1a), typically occupy the ATP-binding cavity in the active "DFG-in" kinase conformation. Type-II inhibitors, such as Imatinib (Figure 1b), not only occupy the ATP-binding cavity but also extend into the adjacent allosteric pockets opened up in the inactive "DFG-out" kinase conformation[20, 35]. Type-III kinase inhibitors, such as Trametinib (Figure 1c), are accommodated just in an allosteric pocket adjoining the ATP binding site. In contrast to Type-III kinase inhibitors, Type-IV kinase inhibitors occupy the allosteric pockets away from the ATP-binding pocket, for example, the allosteric pocket located at the C-lobe (Figure 1d) [36]. By analyzing all available PDB structures, Yueh et. al. identified 10 promising hot spots that are not within the ATP binding site but distributed on the protein surface, as potential binding pockets for designing Type-IV allosteric inhibitors in some kinases[37].

Type I-IV kinase inhibitors can be classified into covalent kinase inhibitors and non-covalent kinase inhibitors based on the presence/absence of kinase-ligand covalent interactions. For example, Crizotinib, Imatinib, Trametinib, and GNF-2 are non-covalent kinase inhibitors (Figure 1a-d). Whereas, Osimertinib is a covalent EGFR kinase inhibitor due to forming an irreversible covalent bond with Cys797 (Figure 1e).

Typically, covalent kinase inhibitors are designed by combining chemical scaffolds with warheads that participate in covalent reactions[21]. The chemical scaffold is generally a proven non-covalent kinase inhibitor, which binds into the designated binding pocket and then provides the foundation[38] for appending a warhead to bear the covalent interactions with proximal nucleophilic residues such as cysteine[39], lysine[40], or tyrosine[41], near or within the binding sites[7, 42-43]. Cysteine is targeted much more frequently than other (non-catalytic) amino acids in covalent drug development, due to its high intrinsic nucleophilicity[7, 10, 44]. To date, eight covalent kinase drugs have been approved by the FDA[21-22]. Nonetheless, owing to potential safety concerns, as a result of protein covalent modification, the pharmaceutical industry remains wary of developing covalent drugs[22]. Accordingly, reversible-covalent inhibition strategies to target protein kinases have been developed[24, 45-46]. The strategies not only avoid permanent protein modification as found with irreversible covalent inhibitors but can also substantially prolong residence time suggesting superior efficacy[28]. In 2012, Taunton and his colleagues reported the first reversible covalent kinase inhibitor by designing and tweaking the reactivity of the warhead moiety of a given irreversible covalent kinase inhibitor, FMK[27]. Since then, reversible-covalent drug design strategies have been applied to multiple kinase targets and numerous RCKIs are reported. Moreover, a promising RCKI, Rilzabrutinib (formerly known as PRN1008), is in a Phase-III clinical trial for Pemphigus treatment[32].

## 2.2 Reaction mechanisms

Generally, reaction mechanisms of covalent inhibitors can be described in the following process[22, 47] :

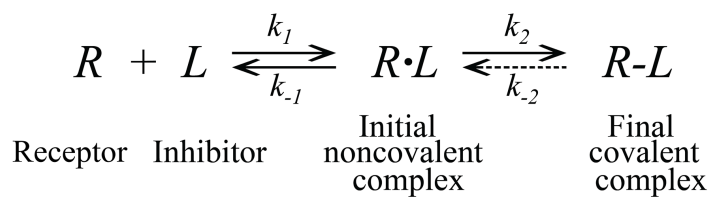

$$R + L \underset{k_{-1}}{\overset{k_1}{\rightleftharpoons}} R \cdot L \underset{k_{-2}}{\overset{k_2}{\rightleftharpoons}} R\text{-}L$$

Receptor   Inhibitor   Initial noncovalent complex   Final covalent complex

where inhibition is a two-step process. First, the inhibitor binds into the receptor binding site forming an initial noncovalent complex. Here, the receptor $[R]$ is in a state of dynamic equilibrium with the inhibitor $[L]$ and the noncovalent complex $[R \cdot L]$, expressed by the noncovalent binding constant ($K_i$), i.e., $K_i = \frac{k_{-1}}{k_1}$. In the second step the receptor is covalently modified and inactivated by covalent-bond formation of the electrophilic warhead of the inhibitor with the adjacent nucleophilic amino acid of the receptor, leading to the final covalent complex. This step is characterized kinetically using a rate constant of protein inactivation $k_{inact}$ (i.e., $k_2$ in the equation above). Thus, the overall covalent binding process is expressed as $k_{inact}/K_i$, a value used to assess the potential of irreversible inhibitor[22-23]. Importantly, conventional IC$_{50}$ values are not ideal to measure the potential of irreversible inhibitors since they are time-dependent[22, 48]. The $k_{-2}$ is the rate constant for covalent dissociation indicating the reversibility for reversible covalent reactions. Copeland et. al.[49-50] first qualified the residence time ($t$) of the drug molecule within its binding site as the reciprocal of the off rate ($k_{-2}$), namely $t = (k_{-2})^{-1}$, for a binary complex model. Studies have shown RCKIs can be designed by embedding the covalent inhibitor into the kinase binding site and tailoring the chemical nature of their warheads to obtain the desirable pharmacodynamics and efficacy[28, 51].

## 2.3 Current status

We manually cataloged all RCKIs published since 2012 using the scholarly literature databases PubMed and Google Scholar. To date, 64 representative RCKIs, inhibiting 10 different kinase targets have been collected (see Table S1 in supporting information). There are eight types of electrophiles (**1 - 8**) (Figure 2) as warheads reported to exhibit reversible-covalent interactions.[44] We review the reaction scheme for each warhead with its respective nucleophile on the kinase protein, i.e., the reversible covalent binding mechanism (Figure 3). Cyanoacrylamide (**1**) is the most frequently used covalent-reversible warhead. Currently, 37 representative RCKIs with warhead **1** have been developed (Table S1). Generally, the reversible covalent reaction of cyanoacrylamide is Michael addition-based and follows a direct reverse process (Figure 3)[52]. Compared with the irreversible warhead acrylamide, cyanoacrylamide has one more nitrile group on the α-carbon. The nitrile is an electron-withdrawing group, inducing the charge redistribution such that the negative charge density around the olefin decreases[27]. The decreased negative charge density not only improves the susceptibility of the β-carbon to nucleophilic attack, but also increases the acidity of the proton on the α-carbon, which facilitates the deprotonation and elimination of the thiol group through the E1cB mechanism, promoting the reversibility of cyanoacrylamide[53]. Chlorofluoroacetamide (CFA, **2**) as the electrophilic group first undergoes an S$_N$2 electrophilic displacement with the thiol group on the protein (Figure 3). However, the reverse reaction of the CFA-thiol adduct undergoes hydrolysis under neutral aqueous conditions (Figure 3)[54]. Mechanically, the S$_N$2 displacement step forms a thiol-substituted-fluoroacetamide intermediate (**2-IM,** Figure 3), where the increased negative charge from the thiol group is redistributed to the α carbon and then to the fluorine atom. The increased electron cloud on the fluorine atom leads to further polarization and easy elimination by forming an F-H bond with the proton during hydrolysis. In contrast, the chloroacetamine-thiol adduct is stable under the same conditions ($t_{1/2}$ > 48h)[54]. Mechanistically, the chloroacetamine-thiol displacement reaction forms a stable C-S bond that is not easily hydrolyzed, unlike the CFA-thiol adduct[54]. Thus, nitrile is a common covalent-reversible warhead[55]. Generally, nitriles exhibit relative inertness, so the formation of covalent adducts requires strong nucleophiles to attack electrophilic carbon atoms. In practice, the electrophilicity of nitriles can be improved by attaching various electron-withdrawing groups[56], such as alkylamines in cyanamide (**3**), and heteroaryl rings in warhead **7**. In the cyanamide (**3**), a nitrogen atom directly attaches to the nitrile. Mechanistically, this adjacent nitrogen atom induces the charge redistribution of the nitrile atoms (i.e., -I effect) and thus the positive charge on the carbon atom of the nitrile is amplified, therefore rendering the carbon atom of the nitrile more electrophilic. Moreover, the nitrile may be further activated by hydrogen bonding to amino acids in the binding site. An aldehyde (**4**) is also a very common warhead in proteolytic enzyme inhibitors[57]. However, aldehydes are not often present in drug discovery because the functional group undergoes additional reactions with off-target enzymes, which produces unexpected toxic adducts[58]. Here the aldehyde is used as an RCKI warhead which may reduce toxicity for two reasons. First, RCKIs are generally designed for selectivity and therefore avoid off-targets[59]. Second, is by tuning the warhead such that covalent reactivity is reversible to avoid permanent off-target modifications[34]. Nevertheless, the high intrinsic reactivity, poor metabolic and chemical stability, and serious health risks associated with the metabolism of aldehydes makes the clinical use of aldehydes problematic.[34, 60]. Cyanoacrylate (**5**) is similar to cyanoacrylamide (**1**) and has the same Michael addition processes as warhead **1**. Cyanoacrylate (**5**) was the first warhead used by Taunton and colleagues in 2012 to design RCKIs, leading to various covalent but reversible warheads including cyanoacrylamide (**1**)[27, 29]. Warhead **6** was recently developed in designing pololike kinase (PLK) 1 inhibitors[61] where the formation of Meisenheimer complexes[62] is considered to be a plausible reaction mechanism (Figure 3). The reversible covalent mechanism was supported by NMR and UV-vis experiments, but no further details were provided, such as binding kinetics assays[61]. Warhead **8** was developed based on "iminoboronate chemistry" [63]. Here the lysine ε-amino group is covalently modified based on the formation of stable iminoboronate with 2-formylbenzeneboronic acid. In designing RCKIs, 2-formylbenzeneboronic acid was morphed into new carbonyl boronic-acid molecules with different scaffolds to achieve prolonged residence time [40, 64]. Next, we review every RCKI based on the described kinase targets and warhead types.

## 2.4 RCKIs

**RCKIs of BTK.** Currently, there are 20 reported RCKIs with high potency and tuned residence time inhibiting BTK (Figure 4). In 2015, Bradshaw et al.[28] reported three RCKIs based on the Ibrutinib scaffold. The acrylamide warhead of Ibrutinib was replaced by cyanoacrylamide (**1**) capped with methyl (**9**), isopropyl (**10**), and tert-butyl (**11**) (Figure 4a). With the increased steric size of the alkyl capping group, the residence time is prolonged, such that **11** – the capping group being tert-butyl - has the longest residence time, showing 55% BTK occupancy 20 h after washout (Figure 4a). The **11**-bound BTK crystal structure shows the binding mode of the scaffold to be similar to Ibrutinib (Figure 4b). The covalent bond is formed between the cyanoacrylamide warhead and the thiol of Cys481, located on the front-pocket rim of the ATP binding site[9, 65]. Specifically, the piperidine amide and the tert-butyl capping group both are oriented to shield the proton which is attached to the Cα. This conformation also prevents overlap between the carbonyl π-system and the Cα-H bond, increasing the kinetic and thermodynamic stability. The hydrogen bond between the amide carbonyl of **11** and the backbone NH of Cys481 strengthens the BTK-**11** interaction. Two of the capping-group methyls form hydrophobic contacts with Leu483 and Arg525, respectively. Thus, for **11**, the hydrogen bond interaction combined with the hydrophobic interaction may further stabilize the covalent complex, leading to a prolonged residence time on the target BTK, compared to RCKIs **9** and **10**. Further, Bradshaw et al. identified additional RCKIs (**12-15**) that improve solubility and oral bioavailability (Figure 4c). Therein, the scaffolds were monofluorinated and more flexible linkers were added to link to the pyrazolopyrimidine scaffolds. Correspondingly, different capping groups were added to the cyanoacrylamide warhead. RCKI **12** showed high potency ($IC_{50}$ = 1.4 ± 0.2 nM) and slow dissociation from BTK (t = 22 ± 3 h). The difference between **12** and **13** is that the methyl-pyrrolidine linkers are a pair of enantiomers where the S-configuration in **13** provides greater potency and slower dissociation ($IC_{50}$ = 0.7 ± 0.1 nM; t = 34 ± 5 h). Based on **13**, inhibitors **14** and **15** were synthesized by capping the warhead **1** with polar, branched-alkyl substituents (i.e., morpholine and oxetane in **14** and **15**, respectively) that have strong binding potential and longer residence time ($IC_{50}$ = 3.2 ± 0.3 nM; t = 83 ± 14 h for **14** and $IC_{50}$ = 1.9 ± 0.3 nM; t = 167 ± 21 h for **15**). Mechanistically, morpholine and oxetane are polar and solvent exposed[28], forming hydrogen-bond interactions with the solvent and the side chain of Arg525 and therefore improve the stability of the covalent complex in **14** and **15**. Further experiments verified that inhibitor **15** has higher selectivity than Ibrutinib based on a 254-kinase panel screen. In this evaluation, 1μM **15** induced >90% inhibition, only BTK and BMX, which is highly homologous and has cysteines at the same positions in the binding sites, were inhibited. A pyrazolopyrimidine fluorescent probe (PP-BODIPY)[66] is an irreversible probe that covalently labels BTK Cys481 with high selectivity and has been shown to penetrate cells[66]. In experiments, the occupancy can be calculated as 100% minus the in-gel fluorescence intensity divided by the control value[28]. As such, PP-BODIPY was used to determine the level of BTK target engagement in rat peripheral blood mononuclear cells (PBMCs) at several times after oral dosing with 40 mg kg$^{-1}$ of **15**. In rodent assays, BTK occupancy by **15** revealed that 41% ± 15% of PP-BODIPY-based probe labeling was blocked 24 h after oral dosing. Although the concentration of **15** in plasma fell to 3 ± 3 ng ml$^{-1}$ at 14 h, it showed significant target engagement and slow dissociation from BTK.

Rilzabrutinib (PRN1008, **16** in Figure 4d) from Principia Biopharma is a BTK inhibitor in Phase-III trials to treat immune thrombocytopenia, pemphigus, and other immunologic disorders[67-68]. With the series discussed in the paragraph above, in contrast to Ibrutinib, its scaffold was monofluorinated and warhead **1** was capped by using a polar, branched-alkyl substituent (Figure 4d). This inhibitor has a high potency (IC$_{50}$ = 1.3 ± 0.5 nM) and long residence time (79 ± 2% of BTK occupancy 18 hours after washing in vitro). A further in vitro assay using a 251-kinase panel screen showed that **16** has high selectivity. In November 2020 the FDA granted a fast-track designation to Rilzabrutinib for the treatment of patients with immune thrombocytopenia. Another Principia Biophama-developed reversible covalent BTK inhibitor, PRN473, has the same scaffold structure as **16**, just differing in the warhead (Figure 4d). Here, the warhead is capped with a tert-butyl group, forming the reversible covalent interaction with Cys481, analogous to the BTK inhibitors discussed above. Currently, PRN473 has completed Phase-I clinical trials for the treatment of neutrophil-mediated tissue damage[33, 68].

Besides warhead **1**, Shindo et al.[54] introduced a chlorofluoroacetamide (CFA) (warhead **2**) to develop a BTK-targeted RCKI (**17** in Figure 4e). RCKI **17** was designed based on Ibrutinib and has the same scaffold. The difference is that RCKI **17** has a CFA warhead bearing a cis-4-substituted cyclohexane linker, unlike Ibrutinib, which has an acrylamide warhead with a 3-substituted piperidine linker. RCKI **17** exhibits strong inhibitory activity on in-cell BTK autophosphorylation ($IC_{50}$ = 44 nM). Moreover, in a Ramos cell assay, RCKI **17** maintained an 82% BTK occupancy 12 h after cell washout, which suggests RCKI **17** has a long residence time.

Another kind of warhead used to design BTK-targeted RCKIs is warhead **3** (cyanamide, Figure 4f-g). Schnute et al.[69] observed that the aminopyrimidine of Ibrutinib forms two conserved hydrogen bonds with the BTK hinge residues Glu475 and Met477 (Figure 4g). Consequently, the authors replaced aminopyrimidine with aminopyrazole carboxamide, without affecting the two conserved hydrogen bonds, resulting in an Ibrutinib-like pseudo-bicyclic arrangement[70]. This modification was used to design a series of novel RCKIs with warhead **3** which achieve reversibility (Figure 4f). The addition reaction was carried out between the cyanamide carbon of warhead **3** and the thiol of Cys481 (Figure 4f, **18-23**). Further modification of the scaffolds, such as pyridinyl substitution (**21-23**), did not impact the potency ($IC_{50}$) or residence time (t) (Figure 4f).

As stated, proteolysis targeting chimeras (PROTACs) are receiving more attention as a treatment modality. Typically, PROTACs are composed of three parts: a protein target binder, a linker, and an E3 ubiquitin ligase ligand. Upon binding, PROTAC forms a ternary complex with the target protein and E3 ubiquitin ligase, leading to ubiquitination and proteasomal degradation of the target protein[71]. Efficient degradation requires strong binding to the protein target. Introducing covalent interactions between the PROTAC and the protein target is one way to strengthen the binding affinity[71]. Recently, several successful irreversible covalent PROTAC degraders have been designed.[71-72] However, irreversible binding may negate the catalytic properties of PROTAC, reducing PROTAC's potency[71-72]. Gabizon[71] and Guo et al.[73] recently reported multiple reversible covalent BTK PROTACs (**24-27**) with high target occupancy and effectiveness as a degrader (Figure 4h). All four PROTAC degraders are Ibrutinib scaffold-based and the electrophile is cyanoacrylamide (**1**). The different linkers capping cyanoacrylamide (**1**) yield the different $IC_{50}s$ and $DC_{50}s$ (i.e., compound concentration inducing 50% protein degradation, Figure 4i). PROTAC dissociation times of 10−20 h are similar to the BTK RCKIs with the warhead cyanoacrylamide[28], which may lead to reduced catalytic efficiency compared to rapid degradation. Importantly, reversible covalent PROTACs maintain the strong covalent interaction with the kinase BTK like RCKI, and significantly improve the selectivity. Therefore, reversible covalent PROTAC degraders are promising, especially for the degradation of a target which has no high-affinity reversible ligands available, and targets where the selectivity of their reversible covalent inhibitors need to be improved[71].

*RCKIs of EGFR.* Rauh's lab reported a series of EGFR-targeted RCKIs to address EGFR drug resistance (Figure 5)[74]. These RCKIs **28-32** (Figure 5b) were designed based on the co-crystal structure of EGFR and pyrazolopyrimidine-based inhibitors published by the same lab. (Figure 5a)[75]. The acrylamide warhead was replaced by cyanoacrylamide (warhead **1**), which forms a covalent bond with Cys797 located at the front pocket of the rim of the binding site[21] (Figure 5a). The different R substituent groups affect the potency of the corresponding inhibitors (Figure 5c). Inhibitor **31** is a strong inhibitor and is over 5-fold more selective for the L858R/T790M mutant ($IC_{50}$ = 20±13nM) over wildtype EGFR ($IC_{50}$ = 96±26nM). Thus, inhibitor **31** can be used as a promising starting point for developing more selective mutant EGFR inhibitors. The authors used a mass-spectroscopy method to characterize the reversible features of RCKIs. This method includes a three-step protocol to characterize RCKIs. First, the apo kinase results in a single peak with a defined m/z value. Second, treatment with an RCKI leads to a characteristic shift of the peak by the molar mass of the RCKI (Peak+$\Delta M_{RCKI}$). Third, after incubation with another covalent inhibitor, another shift of the peak by the molar mass of the covalent inhibitor will occur, with the Peak+$\Delta M_{RCKI}$ disappearing if the RCKI tested is reversible in the timescale of the assay[74].

*RCKIs of JAK3.* The Janus kinase (JAK) family, comprising JAK1, JAK2, JAK3, and TYK2, are attractive targets in the development of anti-inflammatory drugs. Achieving selectivity among JAK family members is an essential yet challenging step in drug discovery resulting from the high degree of structural similarity[76]. The kinase cysteinome[9, 39] reveals JAK3 as having a non-catalytic cysteine residue, Cys909, at the rim of the ATP binding site which does not exist in other JAKs, namely JAK1, JAK2, and TYK2. Over the entire human kinome, MAP2K7, TEC, TXK, ITK, BTK, BLK, HER2, EGFR, and HER4 kinases also contain a cysteine amino acid in the same position as Cys909 of JAK3.[39] Thus, targeting this cysteine is a very promising strategy for achieving isoform-selectivity for JAK3. Forster et al. reported two JAK3-specific RCKIs (**33** and **34**) that bind to the ATP-binding pocket and induce a small binding cleft in the area of the front pocket (Figure 6a-b)[77]. Inhibitors **33** and **34** with $IC_{50}$ values in the picomolar range (127 pM and 154 pM, respectively) show high selectivity (Figure 6a). A binding kinetics assay shows that inhibitor **33** has a prolonged residence time of 50 mins for JAK3. The authors determined the crystal structure of JAK3 kinase domain in complex with **34,** showing the coexistent binding modes of **34** covalently and the non-covalently bound to JAK3 (Figure 6b). Both inhibitors, **33** and **34,** use warhead **1** as an electrophile to form reversible covalent interactions with Cys909. The interactions between the carbonitrile function groups of inhibitors **33** and **34** and nearby residues Arg911, Asp912, and Arg953 induce the aforementioned shallow pocket (Figure 6b), contributing to the selectivity of **33** and **34**. In 2014, London et al.[30] reported a potential RCKI (**35** in Figure 6c) using warhead **1** which covalently targets Cys909 ($IC_{50}$ = 49 nM) and was shown to be reversible based on a dilution experiment[30]. Inhibitor **35** has multiple off-targets, such as BLK ($IC_{50}$ = 22 nM) and HER4 ($IC_{50}$ = 44 nM), and has to be further improved.

Casimiro-Garcia et. al. [78] identified a JAK3 RCKI (**36** in Figure 6d) using warhead **3** (cyanamide) which covalently targets Cys909 ($IC_{50}$ = 456 nM, >22-fold selectivity vs JAK1) and was shown to be reversible (residence time = 154 mins). There are two hydrogen-bond interactions between the pyrrolopyrimidine of inhibitor **36** and the residues Glu903 and Leu905 at the hinge region of JAK3 (Figure 6e). The covalent bond occurs between the nitrile moiety and Cys909 to give an isothiourea adduct (Figure 6e), similar to the binding mode reported for the aforementioned BTK RCKIs (see inhibitor **18**)[69].

*RCKIs of FGFR1*. The FGFR kinases, FGFR-1, -2, -3, and -4, are promising therapeutic targets for multiple types of cancer[79]. Bradshaw et al.[28] reported a series of FGFR1-targeted RCKIs **37-47** (Figure 7a) based on warhead **1** and a pyrimidopyridine scaffold previously used to design irreversible FGFR inhibitors[80]. Here, Cys486, located within the P loop, was used as the nucleophile to form the reversible covalent interaction with the cyanoacrylamide derivative-based warheads in RCKIs **37-47**. RCKIs **37-47** are strong inhibitors ($IC_{50} \leqslant$ 6 nM based on enzyme-activity assays). The different capping groups ($R_1$, see inhibitors **37-47**) on the cyanoacrylamide warhead allowed for the tuning of residence times with the change of 24-hour occupancy rate (%) ranging from 0±10 to 99±8. As such, the range of residence times was from 11 ± 11h to >150 h, highlighting again that simple modifications to the capping group change residence time, which is important for designing drugs with the desired durability. Moreover, the capping groups, provide different residence times on FGFR1 compared to BTK suggesting that the unique protein environments around the cysteine residues also contribute to determining residence time and hence durability[28]. Conceptually, the capping groups, attached to the β-carbon of reversible covalent warheads, form non-covalent interactions with the protein targets near the new covalent bond, resulting in conformational stabilization and masking of the proton on the α-carbon, therefore stabilizing the complex, leading to prolonged residence times[28]. Thus, modifying the capping group has become an effective strategy in designing inhibitors with a variety of residence times[51, 71].

**RCKIs of FGFR4.** FGFR4 is a driver of some solid tumors, for example, rhabdomyosarcoma (RMS) and hepatocellular carcinoma (HCC), and hence has attracted efforts to seek highly selective inhibitors[81]. In FGFR4, a cysteine (Cys552) is located at the GK+2 position of the hinge region[39]. In the entire kinome there are only five kinases with cysteines at the GK+2 position - FGFR4, TTK, MAPKAPK2, MAPKAPK3, and P70S56Kb. FGFR4 is the only FGFR kinase with this cysteine in the FGFR family. Thus, taking advantage of Cys552 would provide selectivity of FGFR4 over FGFR1-3[65]. In doing so, Fairhurst et al. reported highly selective FGFR4-targeted RCKIs (Figure 7b-c)[34, 82-83]. Warhead **4,** an aldehyde**,** was used as the reversible covalent reactive moiety (Figure 2)**.** Based on high-throughput screening and scaffold morphing, inhibitor **56** (also known as Roblitinib or FGF401) was identified as an excellent FGFR4-targeted RCKI and currently is in Phase-II clinical trial for treating HCC and other solid tumors characterized by positive FGFR4 and KLB expression. The Roblitinib-bound FGFR4 complex structure[34] shows inhibitor **56** binds in the ATP-binding cavity and the aldehyde moiety forms a hemithioacetal C-S bond with residue Cys552 (Figure 7b), which greatly improves the selectivity across the whole kinome. In a kinome-screening panel of 456 kinases, it is more than 1000-fold selective for FGFR4 over all other 455 kinases. Other than the strong potency ($K_d$ = 2.9 nM), inhibitor **56** also has a reversible, slow-off inhibitory mechanism (residence time = 52 mins). To get to inhibitor **56**, the authors reported optimizing tens of RCKIs to achieve this best-profile candidate. RCKIs **48-56** represent different structure-activity relationships with corresponding residence times (Figure 7c). Inhibitor 2-FQA (2-formylquinoline amide) was the initial potential FGFR4 covalent inhibitor having been reported by Fairhurst et al. in a previous study[82]. Using inhibitor 2-FQA as a starting point, inhibitor **56** was optimized by both scaffold morphing and substitution by different functional groups ($R_1$-$R_3$)[34]. Specifically, a structure-activity relationship (SAR) analysis surrounding 2-FQA shows its poor solubility and thus the quinoline amide moiety was replaced by a corresponding 2-formyl tetrahydronaphthyridine urea (2-FTHNU) that produced an analogue inhibitor **48** with a modest increased solubility[83]. Surrounding inhibitor **48**, a series of analogues were obtained, such as inhibitors **49** and **50** with similar $IC_{50}$s (28nM (**49**) and 35nM (**50**)) and with 1.5−4.0 h residence times. Further optimization of the 2-FTHNU series was carried out to address the metabolic instability of the warhead aldehyde and hence guarantee robust efficacy in disease-relevant cell lines thereby achieving the desired disease-targeted oral exposure profile[34]. Specifically, a series of 2-FTHNU-based compounds were investigated through $R_1$ substituent SARs and $R_2$ substituent SARs (Figure 7c). The $R_1$-substituted compounds showed that the $R_1$ substituent plays a role in stabilizing the aldehyde group. The introduction of a larger $R_1$ substituent produced 2-FTHNU-based compounds with more favorable metabolic stability. Moreover, of these 2-FTHNU-based compounds, inhibitor **51** exhibited a low-dose oral pharmacokinetic (PK) profile in mouse meaning pharmacodynamic (PD) modulation could be achieved within the acceptable dosage range[34, 84]. Inspired by inhibitor **51**, more promising RCKIs were obtained, such as inhibitor **52** (residence time: 102 mins) and inhibitor **53** (residence time: 107 mins). Inhibitor **54** also showed more favorable low-dose PK profiles in mouse and dog. However, the thermodynamic solubility of inhibitor **54** was still low. Therefore, based upon inhibitor **54,** further 2-FTHNU-based compounds were optimized to improve overall solubility. Finally, more potential compounds were synthesized by adding larger hydrophilic groups (inhibitors **55** and **56;** Figure 7b-c)[34].

*RCKIs of RSK2/MSK1*. Based on previous chemical experiments showing that thiols can react with cyanoacrylates with rapid reversibility at physiological pH[52], Serafimova et al. designed RCKIs based on an irreversible covalent RSK2 inhibitor, FMK(a fluoromethylketone-based inhibitor)[27, 85]. The irreversible warhead (fluoromethylketone) of FMK was replaced by cyanoacrylate (warhead **5**) and its derivative to obtain reversible covalent inhibitors **57** and **59**, and replaced by cyanoacrylamide (warhead **1**) to obtain reversible covalent inhibitor **58** (Figure 8a). The co-crystal structure with inhibitor **59** shows that the thiol group of Cys436 on the β2 sheet forms the covalent C-S bond with warhead **5** (cyanoacrylate). Further experiments verified dissociation of the covalent bonds, as evidenced by unfolding or proteolysis[27] showing the reversible nature of **57-59**. This pioneering work[27] demonstrates the introduction of the reversible warhead cyanoacrylate for designing the first RCKI, and further development of the warhead cyanoacrylamide. Moreover, the authors[27] concluded that the reversible thiol-addition and elimination chemistry seemed to be a general characteristic of cyanoacrylamide, which is very important for a reversible covalent targeting strategy applied to all kinases.

In a separate study, Taunton et al.[29] used an electrophilic fragment-based design strategy to develop RCKIs **60** and **61** (Figure 8b). Inhibitor **60** targets wild-type and T493M RSK2 with a strong potency ($IC_{50}$ = 15 ± 2 nM and 3 ± 1 nM, separately) and was designed using a trimethoxyphenyl-substituted indazole as the scaffold and warhead **1** as the electrophilic group. The scaffold of inhibitor **61** is the same as **60**, but warhead **1** was capped using a 1,1-dimethyl-2-hydroxyethyl group, which keeps the strong inhibitory ability against wild-type and T493M RSK2 ($IC_{50}$ = 13 ± 2 nM and < 2.5 nM, separately). The co-crystal structure of **61** and T493M RSK2 shows a covalent bond formed between the electrophilic β-carbon of the cyanoacrylamide (warhead **1**) and the thiol of Cys436 (Figure 8b). Based on an unfolding experiment of the **61**/RSK2 complex using guanidinium·HCl, the reversibility of inhibitor **61** was verified[29]. The authors also tested whether inhibitor **61** could inhibit the MSK1 C-terminal kinase domain through a reversible covalent interaction with Cys440 ($IC_{50}$ = ~100 nM)[29].

Taunton's group and Shoichet's group[30, 86] reported a virtual covalent docking method (http://covalent.docking.org) used to discover reversible covalent chemical probes by screening large virtual libraries of electrophilic small molecules. As expected, a series of cyanoacrylamide inhibitors were predicted to target RSK2 and MSK1, including inhibitor **62** (Figure 8c), which was an RSK2-targeted probe with a strong potency ($IC_{50}$ = 40 nM). The web server and screening scheme provide a valuable resource for the rapid discovery of potential reversible covalent chemical probes[30].

*RCKIs of PLK1*. Overexpression of Polo-like Kinase 1 (PLK1) is a common feature of cancers, such as gastric and breast cancer[87]. Pearson et al.[61] reported a series of PLK1-targeted RCKIs including inhibitor **63** (Figure 9). Inhibitor **63** forms a covalent bond with Cys67, which is located on the β2 sheet close to the P-loop region. A subsequent experiment verified that inhibitor **63** could effectively inhibit the overexpression of PLK1 with high selectivity (IC$_{50}$ = 2.47 ± 1.23 uM). Inhibitor **63** contains the benzothiazole N-oxide scaffold (warhead **6**) and forms a reversible covalent inhibitor as a Meisenheimer complex (MC), which is stable and fully reversible under normal conditions[88]. The suggested covalent mode of action was supported by UV-Vis spectrophotometry and NMR experiments. The electron rearrangement when forming the adduct between the arene carrying three electron-withdrawing groups and a nucleophile is shown in Figure 9. Changing the electron-withdrawing groups modulates electrophilicity and with it the compound's potency such that inhibitor **64** has higher binding potential (IC$_{50}$ = 0.39 ± 0.07 uM) due to the difference in the R1 substituted group (SCF$_3$), which has a stronger potential to interact with the ATP binding site[16].

***RCKIs of eEF-2K***. Elongation factor 2 kinase (eEF-2K) belongs to the atypical group of kinases and plays an important role in maintaining cellular homeostasis and tumor-cell survival and proliferation[89], making eEF-2K a potential target for cancer treatment, notably breast cancer. Devkota et al. reported an RCKI inhibitor **65**, 2,6-diamino-4-(2-fluorophenyl)-4H-thiopyran-3,5-dicarbonitrile, discovered through structure-based virtual screening (Figure 10a)[89]. Kinetic experiments and molecular docking suggest that RCKI **65** undergoes a reversible covalent mechanism of inhibition where the carbonitrile group (warhead **7**) forms a reversible thioimidate adduct with the eEF-2K's Cys146 located on the β2 sheet of the P-loop region of the ATP binding site (Figure 10b).

***RCKIs of ABL***. The BCR-ABL fusion gene is associated with chronic myelogenous leukemia (CML). Even though multiple ABL-targeted drugs have been developed, frequent acquired drug resistance occurs. Lack of available cysteines at the binding site means no cysteine involved covalent ABL inhibitors against BCR. Recently, Quach et al.[40] designed a class of reversible covalent small-molecule inhibitors targeting the catalytic lysine residue of the β3 sheet in the N-lobe of ABL (part of the ATP binding pocket) using an iminoboronate strategy. A strategy used previously to reversibly, but covalently, modify amino groups in proteins[63]. The two inhibitors **66-67** were designed based on a non-covalent pan-ABL inhibitor PPY-A[90] (Figure 11a) equipped with the reversible covalent warhead **8** (aldehyde boronic acid, Figure 2). The co-crystal structure of the inhibitor **66**-bound ABL complex illustrates that the aldehyde (warhead **8**) forms an imine by reacting with the amine of Lys271 in the ATP-binding pocket (Figure 11b). The boron atom of the boric acid functional group and the imine nitrogen of Lys271 form the coordinate bond interaction, suggesting that boric acid plays a key role in the addition process and the final formation of the adduct. The covalent interaction yields high potency: inhibitor **66** with $IC_{50}$ (WT) = $1.7 \pm 0.2$ nM, $IC_{50}$(T315I) = $0.1 \pm 0.1$ nM, and $IC_{50}$ (E255K) = $0.5 \pm 0.03$ nM; and inhibitor **67** with $IC_{50}$(WT) = $5.0 \pm 0.4$ nM. The reversibility of inhibitors **66-67** was confirmed using a $NaBH_3CN$ labeling experiment and an alkyne-containing analog (**68**). $NaBH_3CN$ is a reducing agent that traps the iminoboronates as more stable amines[63]. In the presence of $NaBH_3CN$, an increase in the fluorescence intensity of **68**-labeled proteomic samples was observed. In contrast, washing the **68**-labeled proteome with cold acetone and methanol in the absence of $NaBH_3CN$ significantly reduced or eliminated the fluorescence signals. Thus, the reversibility of the iminoboronate bond was confirmed. The authors also showed further evidence for the reversibility of the iminoboronate bond using $^1H$ NMR studies of the imine product formed between 2-formylphenyl boronic acid and Ac-Lys-NHMe[40].

## 3. RCKIs Design Strategies

### 3.1 Major strategies to develop RCKIs

Based on prior work, three major strategies for developing RCKIs emerge. The first is to tune the electrophilic groups of existing irreversible-covalent kinase inhibitors towards reversibility. Existing covalent kinase inhibitors have a well-defined scaffold and the nucleophilic group has been validated to be within striking distance. For example, Ibrutinib is a prototypical BTK covalent kinase inhibitor. Based on Ibrutinib, many BTK-targeted RCKIs, such as RCKIs **9-11**, have been reported by creating reversible warheads.

The second strategy is to engineer a reversible covalent electrophilic group based on an existing non-covalent kinase inhibitor. For example, JAK3-targeted RCKIs **33** and **34** were developed by using a non-covalent JAK3 compound ($IC_{50}$ = 63 nM, and Figure S1) as the starting point. Specifically, a linker moiety bearing an electrophile was added to the compound to suitably interact with the Cys909 of JAK3[77]. Thus, high potency and selectivity of RCKIs **33** and **34** were obtained and proven in a cellular model[77]. Given the large number of known non-covalent kinase inhibitors, and the increasing repository of kinase structures[16], this non-covalent inhibitor-based strategy is a valuable approach.

The third strategy is a high-throughput virtual screen. In this RCKI screening protocol, a covalent docking method is used to anchor the covalently interacting atoms. Virtual screening extends the chemical conformation space of RCKIs and provides opportunities to design chemically novel RCKIs. For example, London et al. applied a virtual screening scheme for discovering novel RCKIs and AmpC β-lactamase-targeted reversible-covalent inhibitors[30]. However, virtual screening method while easy to use, produce false hits and artificial binding poses, thus further experimental validation is warranted[91].

**3.2 Privileged reversible-covalent warheads**

The electrophile is the centerpiece of designing RCKIs. The nature of the electrophilic group determines the reversibility and the residence time. Currently, there are 8 types of warheads reported (Figure 2). The corresponding derivatives, such as cyanoacrylamide, capped with different substituents contribute to the selectivity and residence time. For example, inhibitors **37- 47** achieve different residence times by modifying warhead **1** using different substituent groups (Figure 7). Therefore, choosing a potential electrophile and further tuning its reversible-covalent electrophilic properties are essential steps. One way to choose an electrophile is from available warhead libraries[92]. Given the success of irreversible covalent kinase drugs, multiple electrophilic warhead databases for irreversible covalent kinase drug design are available[29, 39, 92]. With this warhead toolbox researchers can tweak the characteristics of the chosen irreversible covalent electrophile to establish the reversible covalent interactions. Krishnan et al.[53] used a computational method to analyze the intrinsic reaction trends of different electrophilic groups during the thiol-Michael addition reaction. This intrinsic trend is beneficial in guiding the design of the desired reversible covalent interaction. The other way to choose an electrophile is based on the intrinsic reversibility of chemical reactions between the electrophile and nucleophile. For example, experiments in 1968 found that the chemical reaction between thiol and cyanoacrylamide was reversible[52]. Currently, cyanoacrylamide has been used to successfully design RCKIs to target BTK, EGFR, JAKs, and others. As such, cyanoacrylamide is a privileged reversible-covalent warhead based on the intrinsic reactive properties and proven successful applications.

### 3.3 Nucleophilic residues in the binding site

Just as important, are nucleophilic groups within warhead striking distance[39, 93]. So far, multiple nucleophilic amino acids, including cysteine and lysine, from 10 distinct kinases have been targeted as RCKIs (Figure 11a). The 10 kinases belong to 4 kinase groups (Figure 11a): (i) the TK group including the kinases ABL, and BTK. EGFR, JAK3, FGFR1, and FGFR4; (ii) the AGC group including the kinases MSK1 and RSK2; (iii) the kinase PLK1, and (iv) the atypical group including the kinase eEF-2K. More importantly, the nucleophilic groups are located at different positions within the binding sites (Figure 11b), including the cysteine (Cys552) at the Hinge region targeted by the RCKIs of FGFR4; the cysteine (Cys486) at the P-loop targeted by the RCKIs of FGFR1; the cysteines (Cys481 in BTK, Cys797 in EGFR, and Cys909 in JAK3, respectively) at the Front pocket at the rim of binding site targeted by the RCKIs of BTK, EGFR, and JAK3; the cysteines (Cys436 in RSK1, Cys440 in MSK1, Cys67 in PLK1, and Cys146 in eEF-2K, respectively) at the β2 sheet targeted by the RCKIs of RSK1, MSK1, PLK1, and eEF-2K; and the catalytic lysine (Lys271) at the β3 sheet targeted by the RCKIs of ABL1. Targeting lysine amino group with covalently interacting electrophilic ligands has been a challenging task due to the high pKa (~10) leading to complete protonation under physiological condition (pH=7.4). In this study, RCKIs **66** and **67** were designed and validated as targeting the conserved catalytic lysine of ABL1 (Lys271). These results suggest that designing selective RCKIs to target the conserved lysine in ATP binding sites may be practicable[40, 94]. Based on the previous study of the cysteinome[9, 21, 39], there are over 200 kinases with accessible cysteines within striking distance, providing an abundant structural resource to design RCKIs[9]. Our previous study concluded that the microenvironment of the cysteine inside the binding sites affects the chemical reactivity of covalent interaction[65]. Microenvironment also affects the reversibility of the covalent reaction. For example, Shindo[54] reported a series of RCKIs using a chlorofluoroacetamide (CFA, Warhead **2**) as the electrophilic group and experimentally validated that the CFA-based thiol adduct was easily hydrolyzed under neutral aqueous conditions. Despite the weak intrinsic reactivity of CFA, CFA-appended probes, based on the molecular architecture of Afatinib as the scaffold, showed high reactivity to the Cys797 of EGFR. The linkers connecting the warhead CFA and the scaffold affected the reaction potency[54]. Moreover, the CFA-based EGFR probes were stable in the solvent-sequestered EGFR ATP binding pocket ($t_{1/2} > 72h$)[54]. In contrast, the CFA-based probes showed reversible properties when targeting Cys481 of BTK, such that **17** maintained an 82% BTK occupancy rate 12 h after cell washout, which indicated that Cys481 in BTK is more solvent-accessible than the Cys797 in EGFR[54]. This suggests that the influence of the microenvironment near the nucleophilic residue should be taken into account during designing RCKI warheads, especially for CFA, a solvent environment-sensitive warhead, to target more non-catalytic cysteines located at different positions of the kinase domain[95-96].

## 4. Conclusion and outlook

Here we systematically describe progress with RCKIs, summarizing the different types of reversible covalent warheads and the corresponding striking nucleophilic residues among 10 different kinases, highlighting design strategies for RCKIs, the privileged reversible covalent warhead cyanoacrylamide, and effects of microenvironments near the nucleophilic residues especially for the solvent-sensitive warhead CFA. We summarize three RCKI design pipelines, which have benefited from the current successes with reversible covalent kinase inhibitors and diverse kinase inhibitors. Notably, a large number of active kinase inhibitors, such as the ~54,000 active compounds in ChEMBL Kinase SARfari, provide abundant structure-activity relationships and design opportunities[97-98]. Also, elaborating upon the reversibility of electrophilic groups is an essential step in obtaining reversible covalent characteristics. Multiple research groups have systematically studied the intrinsic reactivity of diverse warheads[53, 92, 99], which helps determine the right electrophilic groups as the starting point for designing RCKIs. Fortunately, warhead **1** showed privileged reversible covalent properties and has been the most frequently used electrophilic group and has been designed to target different kinases including BTK, EGFR, JAK3, RSK2, and MSK1. Correspondingly, abundant cysteine residues, located in different positions of the binding sites in about 200 kinases, and catalytic lysine residues across the whole kinome, can act as nucleophiles, thereby providing tremendous opportunities for developing RCKIs. The reversibility of electrophilic groups should be specifically estimated when the nucleophilic groups are located in different microenvironments, such as solvent-accessible or solvent-sequestered sites.

Measured against the whole human kinome, the field of RCKI development is far from mature. With the anticipated approval of reversible covalent drugs, their advantages, especially the tunable residence time, will attract more attention in reversible covalent drug discovery to treat chronic diseases. For instance, Rilzabrutinib, a Phase-III drug[68, 100], illustrates the potential by utilizing its rapid reversibility avoiding unwanted adverse effects on the immune system. RCKIs may have another advantage with targets having a high turnover, as shown in the development of FGF401, where the rapid turnover of FGFR4 led the Novartis researchers to pursue a reversible-covalent approach instead of an irreversible one[34, 82-83]. In conclusion, we can expect more kinases to be targeted by reversible covalent inhibitors.

## 5. Supporting Information

The complete list of all reported RCKIs and general reaction schemes (docx).

## 6. Data Sharing and Data Availability

The data that supports the findings of this study are available in the supplementary material of this article.

# 7. Acknowledgements

We thank the two anonymous expert reviewers for their professional and insightful comments and suggestions.

**AUTHOR BIOGRAPHIES**

**Zheng Zhao** is a research scientist in the School of Data Science and School of Medicine at the University of Virginia. He holds a B.S. in chemistry from Wuhan University and received a Ph.D. in bioinformatics in 2008 from the University of Science and Technology of China. His research mainly focuses on the development and application of data-driven, computational drug discovery approaches for novel drug design and discovery, drug repurposing, drug mechanisms of action.

**Philip E. Bourne** is the Stephenson Founding Dean of the School of Data Science, a professor of Data Science and a professor of Biomedical Engineering at the University of Virginia. He received his PhD in chemistry from the Flinders University of South Australia in 1980. His research career has spanned structural biology, bioinformatics, systems biology, systems pharmacology and most recently data science more broadly.

**Figures:**

**Figure 1.** Kinase inhibitor binding modes. (a). Type-I inhibitor Crizotinib in a "DFG-in" conformation (PDB id: 3zbf); (b). Type-II inhibitor Imatinib in the "DFG-out" conformation (PDB id: 1opj); (c). Type-III inhibitor Trametinib (PDB id: 7jur); (d). Type-IV allosteric inhibitor GNF-2 bound to the allosteric pocket of the C-lobe (PDB id: 3k5v); (e). Covalent EGFR kinase inhibitor Osimertinib with a covalent bond interaction with Cys797 (PDB id: 6jxt).

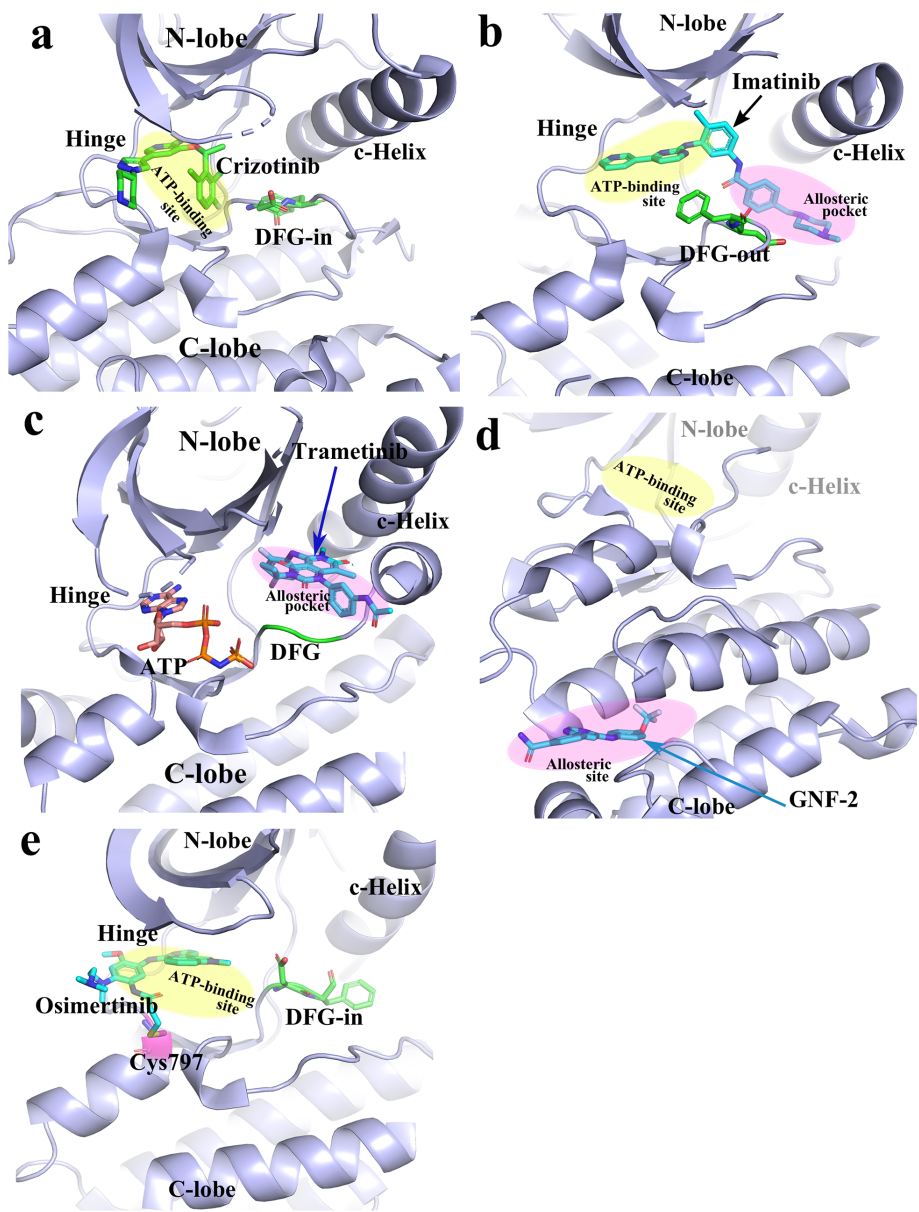

**Figure 2.** Warhead moieties in RCKIs. Stars mark the active atoms in reversible covalent reactions.

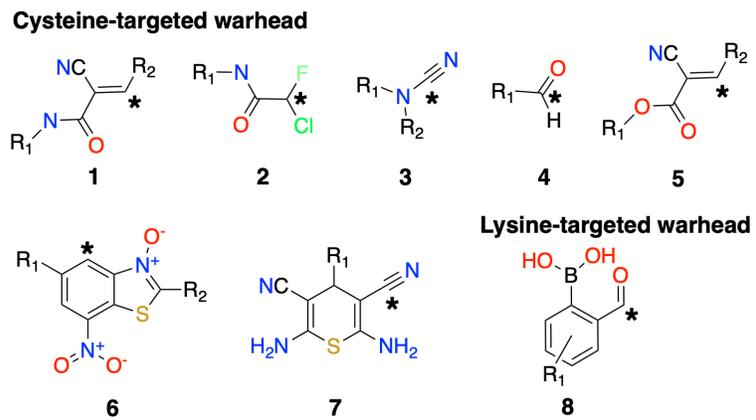

**Figure 3.** General reaction schemes of the warheads in RCKIs to bind to the corresponding nucleophiles.

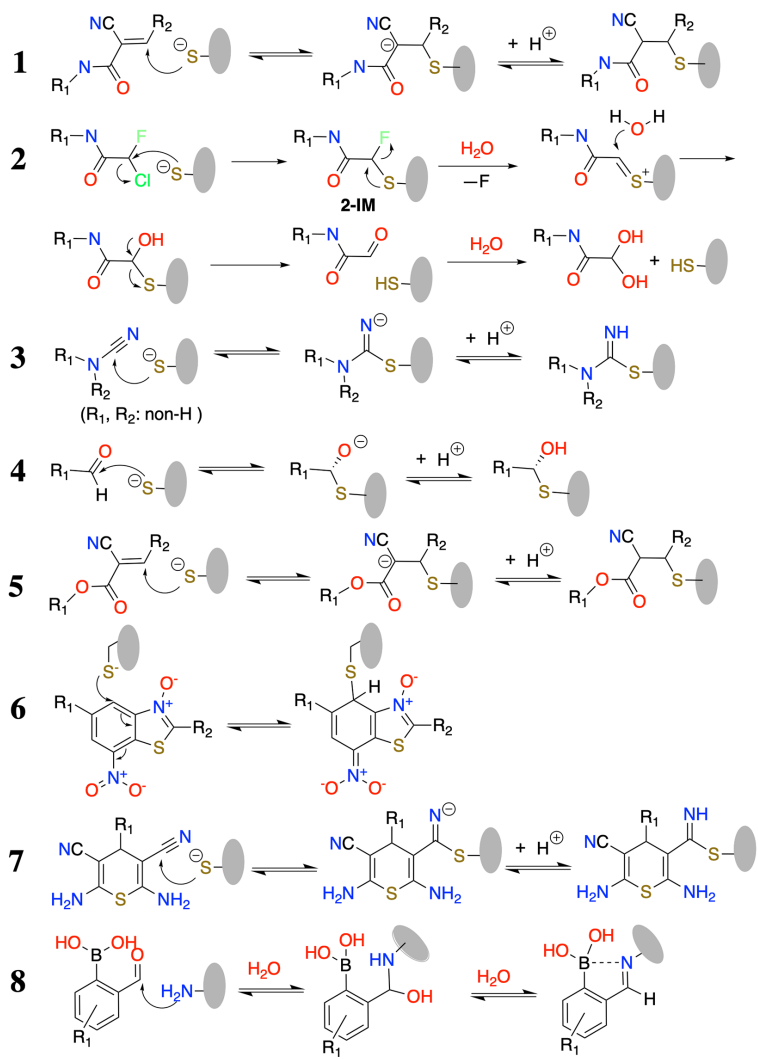

**Figure 4**. 2D structures and characteristics of RCKIs inhibiting BTK. (a) Inhibitors **9-11**, the corresponding inhibitory potencies and the residence time. (b) the binding mode of inhibitor **11** (PDB id: 4yhf, grey) compared to Ibrutinib (PDB id: 5p9i, light-blue). The covalent bond formation between Cys481 and Cβ. (c-e) Inhibitors **12-17**. (f) Inhibitors **18-23**, the corresponding inhibitory potencies and the residence time, and (g) the cocrystal structure of the covalent adduct between compound **18** and the mouse BTK kinase domain (PDB id: 6mny). (h-i) PROTACs **24-27** and the corresponding inhibitory potencies and degradation rates.

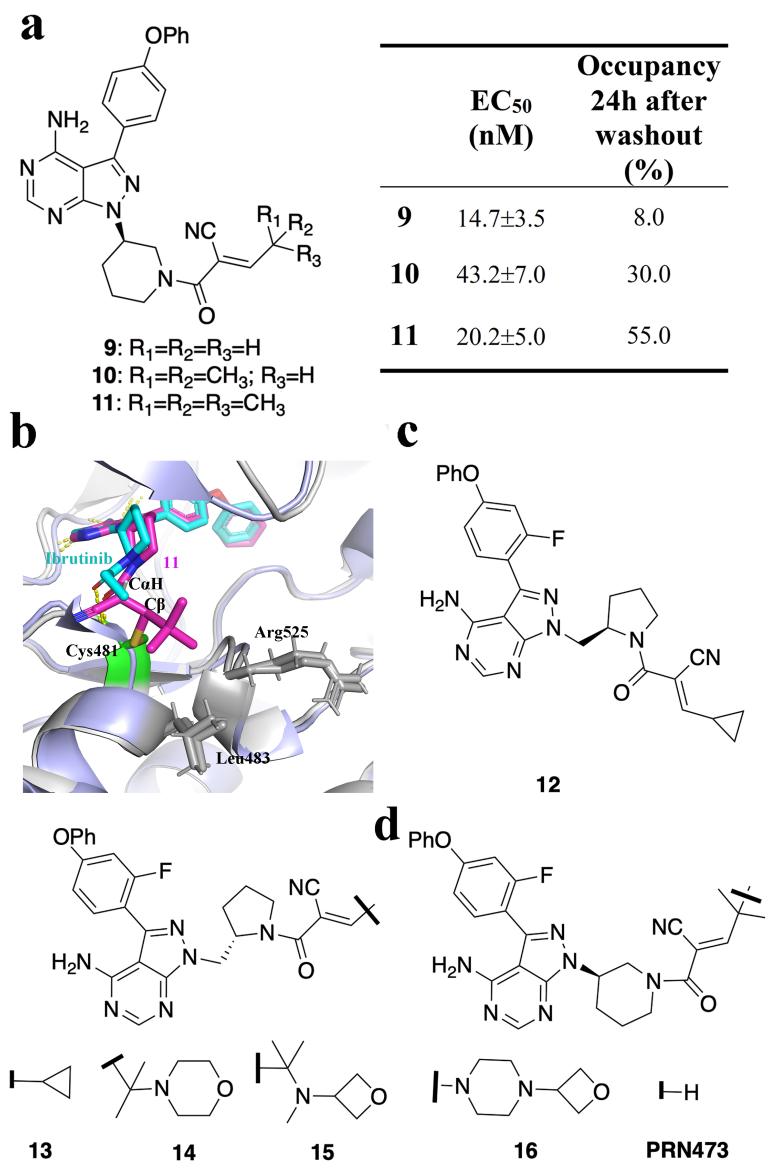

**Figure 4. Continued**

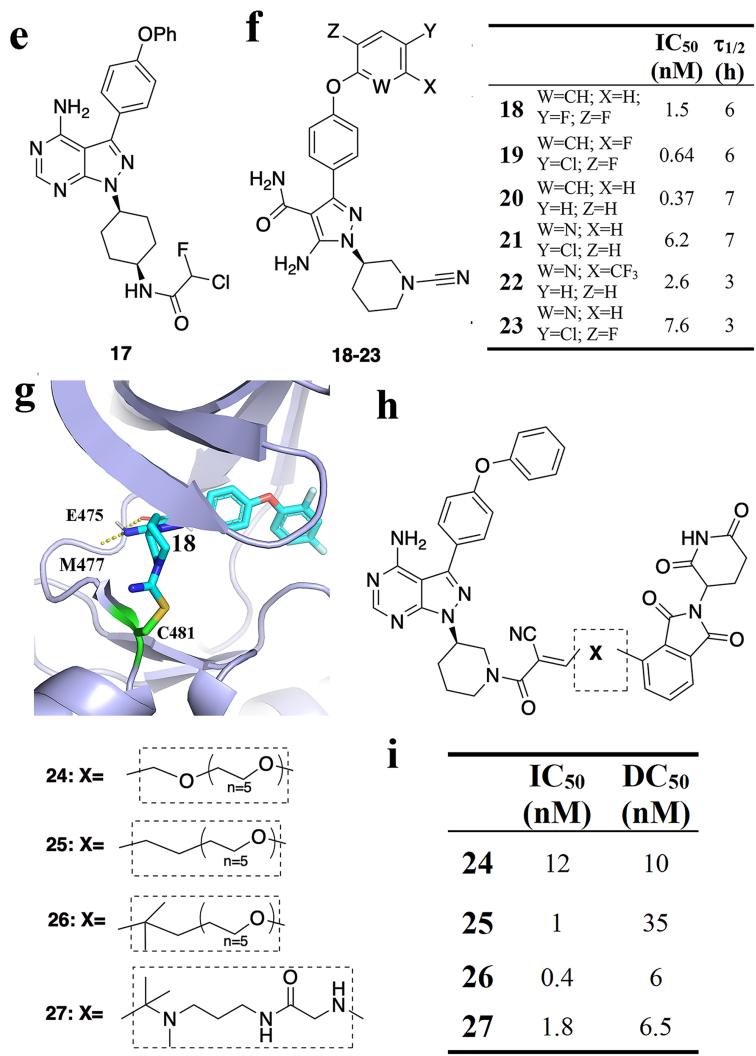

**Figure 5.** (a) Binding mode of the pyrazolopyrimidine-framework covalent inhibitor (PDB id: 5j9y); (b-c) RCKIs of EGFR and the corresponding binding affinities.

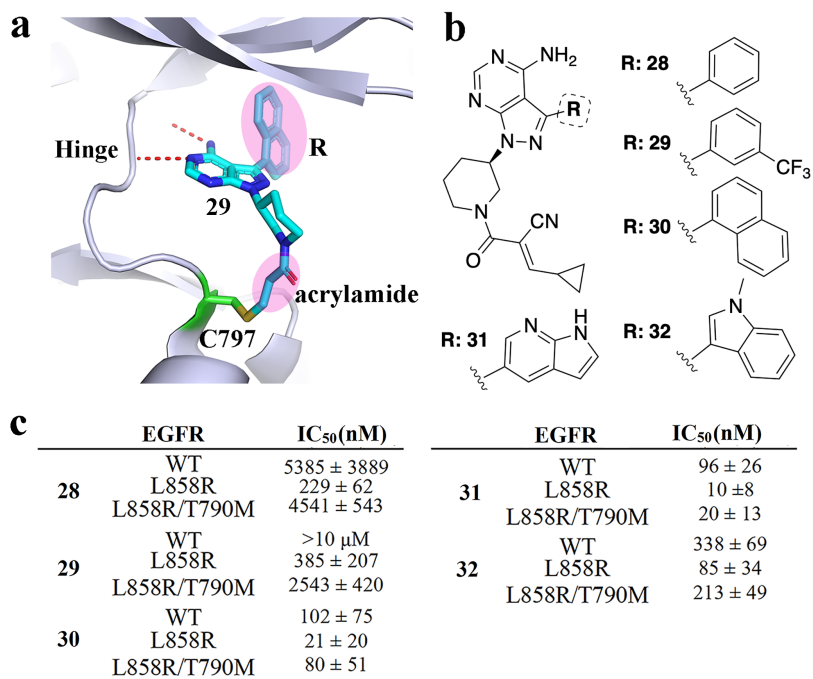

**Figure 6.** RCKIs of JAK3. (a) Chemical structures of inhibitors **33** and **34** and the corresponding inhibitory activity within the JAK family. (b) Coexistent binding modes of the covalently and the noncovalently JAK3-bound inhibitor **34** (PDB id: 5lwn). (c) 2D structure of inhibitor **35** with warhead **1**. (d) 2D structure of inhibitor **36** with warhead **3**. (e) Crystal structure of 36 (green) bound to JAK3 (lightblue). Formation of a covalent bond with Cys909 was observed (PDB id: 6da4).

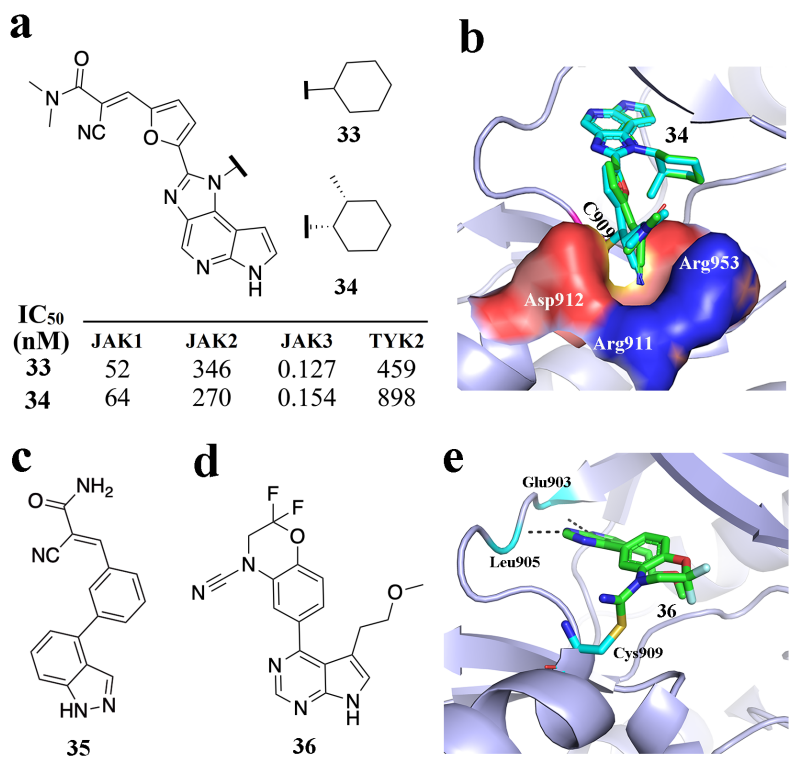

**Figure 7.** (a) 2D chemical structures of RCKIs of FGFR1. Warhead **1** is shown in the dashed rectangle. (b) Binding mode of Roblitinib based on the co-crystal complex structure (PDB id: 6yi8). The yellow dash lines show the H-bond interaction between Roblitinib and the amino acids R483, V500, and A553. (c) The FGFR4-targeted RCKIs with the potency and the corresponding residence time.

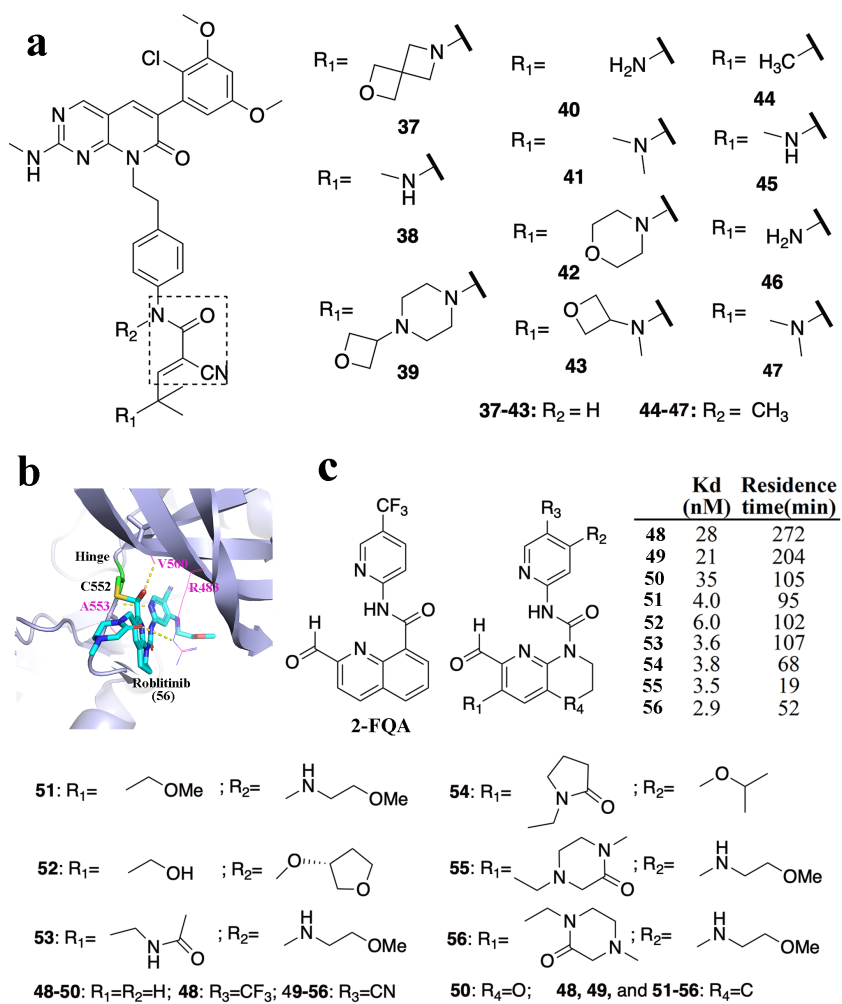

**Figure 8.** (a) 2D chemical structures of RCKIs **57-59** and the binding modes of **59** shown in the co-crystal inhibitor-RSK complexes (PDB id: 4d9u). (b) 2D chemical structures of RCKIs **60-61** and the binding modes of **61** (PDB id: 4jg8). (c) 2D chemical structures of RCKI **62**.

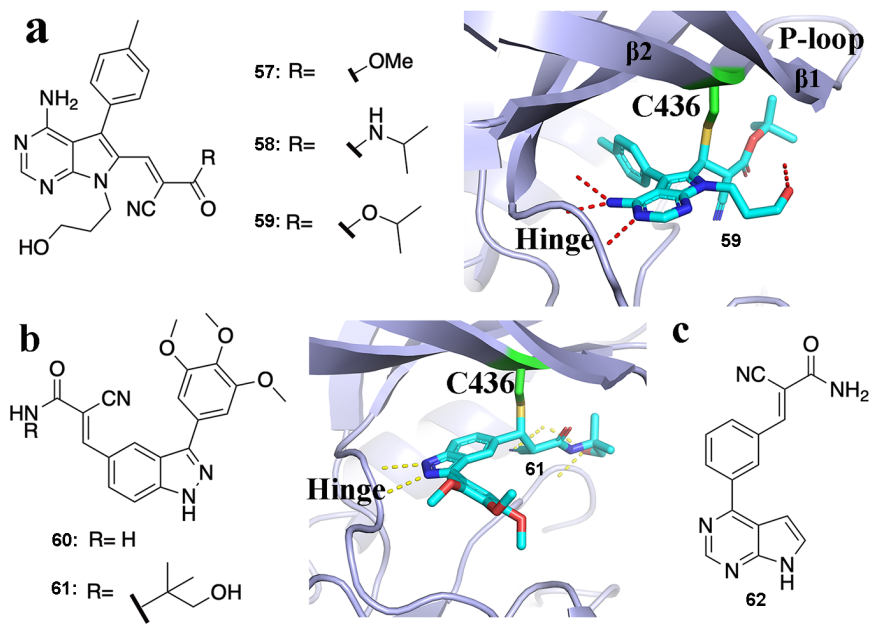

**Figure 9.** Chemical structures of PLK1-targeted RCKIs **63-64** showing benzothiazole N-Oxide Meisenheimer complex formation.

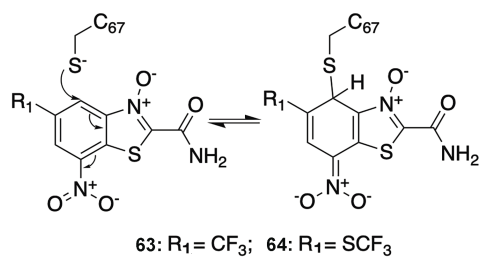

**63**: $R_1$ = $CF_3$;  **64**: $R_1$ = $SCF_3$

**Figure 10.** (a) Chemical structure of eEF-2K-targeted RCKI **65**. (b) Binding mode of inhibitor **65**, adapted from Devkota et al[89].

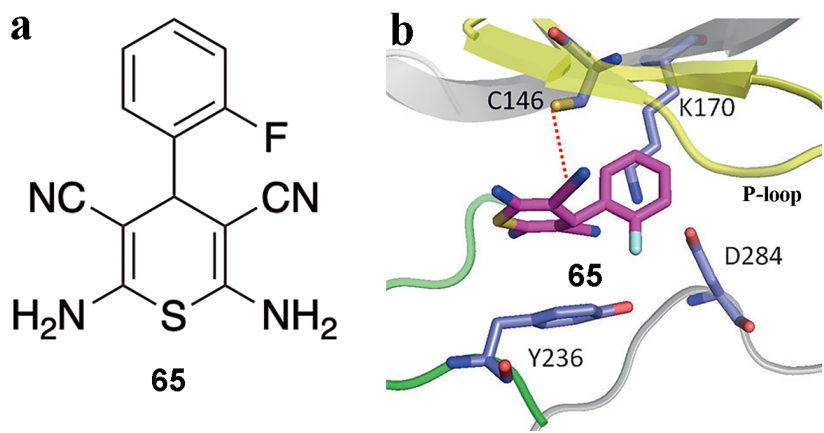

**Figure 11.** (a) 2D structure of PPY-A marked by a dashed rectangle and 2D structures of RCKIs **66-68** of ABL. (b) The binding mode of RCKI **66** (PDB id: 7dt2).

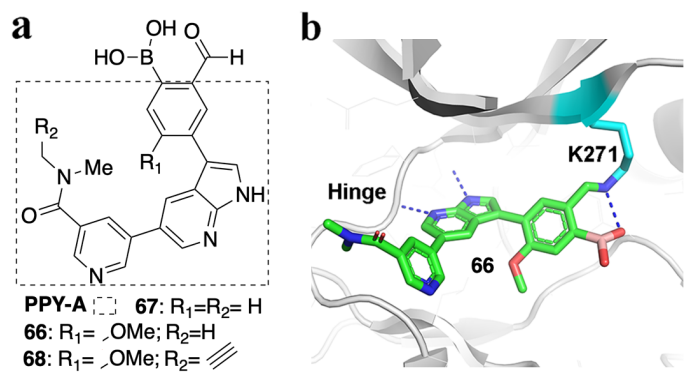

**Figure 12**. (a) Distributions of kinases with the released RCKIs (Image generated using TREEspot™ Software). (b) Distributions of nucleophilic amino acids near the ATP binding sites in different kinases. Here, only nucleophilic positions that have already been targeted by RCKIs are shown. The different colored balls show diverse types of nucleophilic residues. Purple, cysteine; Blue, lysine.

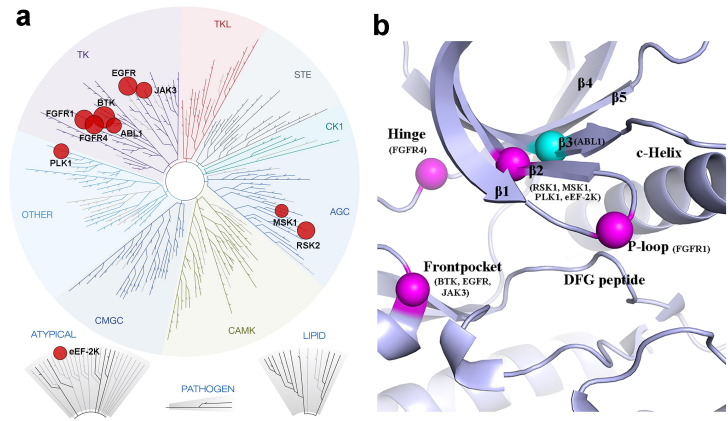